\documentclass[fleqn,10pt]{wlscirep}
\usepackage{graphicx}
\usepackage{amsmath}

\title{The scaling of income inequality in cities}

\author[1,*]{Somwrita Sarkar}
\author[1]{Peter Phibbs}
\author[1]{Roderick Simpson}
\author[1]{Sachin N. Wasnik}
\affil[1]{Faculty of Architecture, Design, and Planning, University of Sydney, NSW 2006, Australia.}

\affil[*]{somwrita.sarkar@sydney.edu.au}

\begin{abstract}
Developing a scientific understanding of cities in a fast urbanizing world is essential for planning sustainable urban systems. Recently, it was shown that income and wealth creation follow increasing returns, scaling superlinearly with city size. We study scaling of per capita incomes for separate census defined income categories against population size for the whole of Australia. Across several urban area definitions, we find that lowest incomes grow just linearly or sublinearly ($\beta = 0.94$ to $1.00$), whereas highest incomes grow superlinearly ($\beta = 1.00$ to $1.21$), with total income just superlinear ($\beta = 1.03$ to $1.05$). These findings support the earlier finding: the bigger the city, the richer the city. But, we also see an emergent metric of inequality: the larger the population size and densities of a city, higher incomes grow more quickly than lower, suggesting a disproportionate agglomeration of incomes in the highest income categories in big cities. Because there are many more people on lower incomes that scale sublinearly as compared to the highest that scale superlinearly, these findings suggest a scaling of inequality: the larger the population, the greater the inequality. Urban and economic planning will need to examine ways in which larger cities can be made more equitable. 

\end{abstract}
\begin{document}

\flushbottom
\maketitle
%
%
\thispagestyle{empty}


\section*{Introduction}

As the world urbanizes itself faster than ever, it is becoming more and more important to understand the principles behind urban spatial and socio-economic structure, in the hope that we will be able to design and plan cities better. Cities are very special examples of complex systems that are \textit{both} self-organized and designed at the same time. This separates them from large classes of other systems that are primarily either completely self-organized (e.g. biological networks), or completely designed (e.g., complex engineering systems such as airplanes or spaceships). In cities, centralized planning and policy making co-exists with distributed responses to plans and policies from millions of inhabitants, and together these produce the spatial and socio-economic structure of the city~\cite{batty2013, barthelemy2013}. Despite decades of research, a quantitatively based theory of cities is still missing. 

Recently, it was proposed that one possible way of quantitatively characterizing urban spatial and socio-economic structure is through universal scaling laws of the form~\cite{bettencourt2007, bettencourt2010, bettencourt2010a} \begin{equation} Y(t) = Y_{0}N(t)^{\beta}, \end{equation} where $N(t)$ is the population at time $t$, and $Y$ denotes material resources (such as energy, infrastructure or dwelling stock) or measures of social and economic activity (including both advantages such as incomes, patents, research and development activity, or disadvantages such as crime and pollution). With $Y_{0}$ as a normalization constant, the scaling exponent $\beta$ reveals the dynamics of whether an urban indicator or measured quantity scales super-linearly, linearly, or sub-linearly depending on whether the value of $\beta$ is estimated to be below, at, or above 1, respectively. 

Studies were performed across several Metropolitan Statistical Regions (MSAs) in the USA, and other cities in China and Europe. It was found that all material resources (such as infrastructure, road networks) showed economies of scale and scaled sub-linearly with population size, with the value of $\beta$ less than 1 or almost 1 (such as dwelling stock). On the other hand, most social and economic indicators (such as incomes, wealth, patents, crime) showed increasing returns and scaled superlinearly with population size, with the value of $\beta$ consistently more than 1. This leads to the postulated theory behind the existence of cities: urban agglomerations exist because it is inherently advantageous for them to exist. As population grows, the per capita expenditure on maintaining the urban system is less than the per capita socio-economic returns by way of income generation and wealth (although negative aspects such as crime grow superlinearly too).

This previous analysis was carried out over aggregated measures such as total wages, total numbers of patents, or total bank deposits. However, while cities are wealth and knowledge creators, they also demonstrate heterogeneities, inequalities of resource distributions, and polarizations of social and economic indicators. In particular, the issue of income and economic inequality has been in focus, historically~\cite{sen1973, sen1997}, as well recently as witnessed through several timely publications and wide ranging public and scholarly debates~\cite{wilkinson2009, stiglitz2012, piketty2014, atkinson2015}. In this paper we focus on the question: Does income inequality scale with city or population size? Because this is an initial examination, we focus only on income as reported in the census, but in future work, the approach presented here can be extended to more in-depth and derived measures of income, such as computed disposable incomes, incomes in relationship with social need or functions, or other measures of economic well being or equity~\cite{sen1997, atkinson2015}. 

We examine the scaling of per capita income against population size for different categories of income earners, across several urban area definitions based on social and economic geography for the country of Australia. Instead of focusing on aggregate measurements of income and wealth in cities, as previous studies have done, we focus on specific income categories: how much income is earned in a specific income category, or equivalently, what is the distribution of people in specific income categories, and how does this measure scale with city or population size? 

There is now considerable debate over the definition of what a city is~\cite{arcaute2015}. Administrative boundaries often do not coincide with economic or social activity, population density, or journey to work patterns, and this has been shown to affect the scaling dynamics~\cite{arcaute2015}. Therefore, in order to test the universality of our results, we perform the analysis over several urban area definitions, based on statistical and economic labor force region definitions provided by the national statistical agency, the Australian Bureau of Statistics (ABS), and using these to generate both population count as well as population density based urban area definitions. The ABS is a unique body in the sense that the geographic area definitions it provides are, to a large extent, defined on the basis of labor markets and population counts and densities in large regions. For more discussion of the finer points of area definitions used for the analysis, see Methods.  

We choose the country of Australia for several reasons. First, while studies of US, Europe, and separately the UK have been reported~\cite{bettencourt2007, arcaute2015}, Australia's unique geographic position as an island-country-continent merits study (though New Zealand is not included in our analysis, due to separate census bodies in the two countries). Second, even though Australia is one of the most urbanized countries in the world, it is also one of the most sparsely populated and the youngest: the urban structure is still nascent. The population in Australia is not spread out as USA, Europe, or UK. The urban landscape is dominated by the Sydney and Melbourne metropolitan regions, with 3-4 smaller regions (Perth, Adelaide, Brisbane), followed by a few other smaller urban areas, and finally a vast, mostly uninhabited continent in the middle. This results in a situation where almost the entire population (more than 90\%) is urban and lives agglomerated in a few large urban areas, with the rest of the continent very sparsely inhabited. Third, because of this unique position, a single national body, the Australian Bureau of Statistics (ABS) maintains data across several spatial scales, where in a large number of ways the administrative boundaries concur reasonably with real economic and social boundaries, population densities, and other bottom up indicators that have recently been shown as important factors affecting the scaling behavior~\cite{arcaute2015}. This allowed us to test the variation and sensitivity of the scaling behavior across several stable urban area definitions. These special factors render the Australian case study important: how does scaling behavior change with a significantly different geographic prototype such as Australia? 
 

\section*{Results}

The ABS carries out a detailed census once every 5 years. For this paper, we have worked with the latest census data from 2011. The urban area definitions we have used are discussed in detail in the Methods section and are derived using the Australian Statistical Geography Standards (ASGS). 

\subsection*{Geography: Significant Urban Areas (SUA), Population and Density Cutoffs, and Income Categories}

We have used Significant Urban Areas (SUA) to define the urban area geography of Australia (see Methods). There are 101 SUAs defined for Australia. The five largest largest SUAs (continuous urban areas or ``cities") in Australia are Sydney, Melbourne, Perth, Adelaide, and Brisbane, with about 60\% of the population of the entire country residing in a captial city, and about 35\% in Sydney and Melbourne alone. These centres have populations greater than 1 million. We have performed the scaling analysis firstly for all 101 SUAs, followed by considering alternate urban area definitions. Several smaller subsets of the SUAs are considered by including high population density and high population counts as cutoffs, and excluding very low population SUAs and very low population density SUAs (especially those that are of an overwhelmingly regional character surrounded by large rural hinterlands), and measured how the scaling exponent behavior changes under urban area defintions that look at increasing population density and total counts. 

For each statistical area, the ABS has defined 10 income categories and provides the imputed median incomes for these categories (see Methods). In addition, the census of 2011 provides the count of the number of people in each of these income categories per SUA. We use the ABS imputed median incomes and count of people in each category to produce the \textit{computed median incomes} for each SUA (see Methods).

\subsection*{Scaling of income in different income categories: All SUAs}

\begin{figure}[]
\centering
\includegraphics[width=0.8\linewidth]{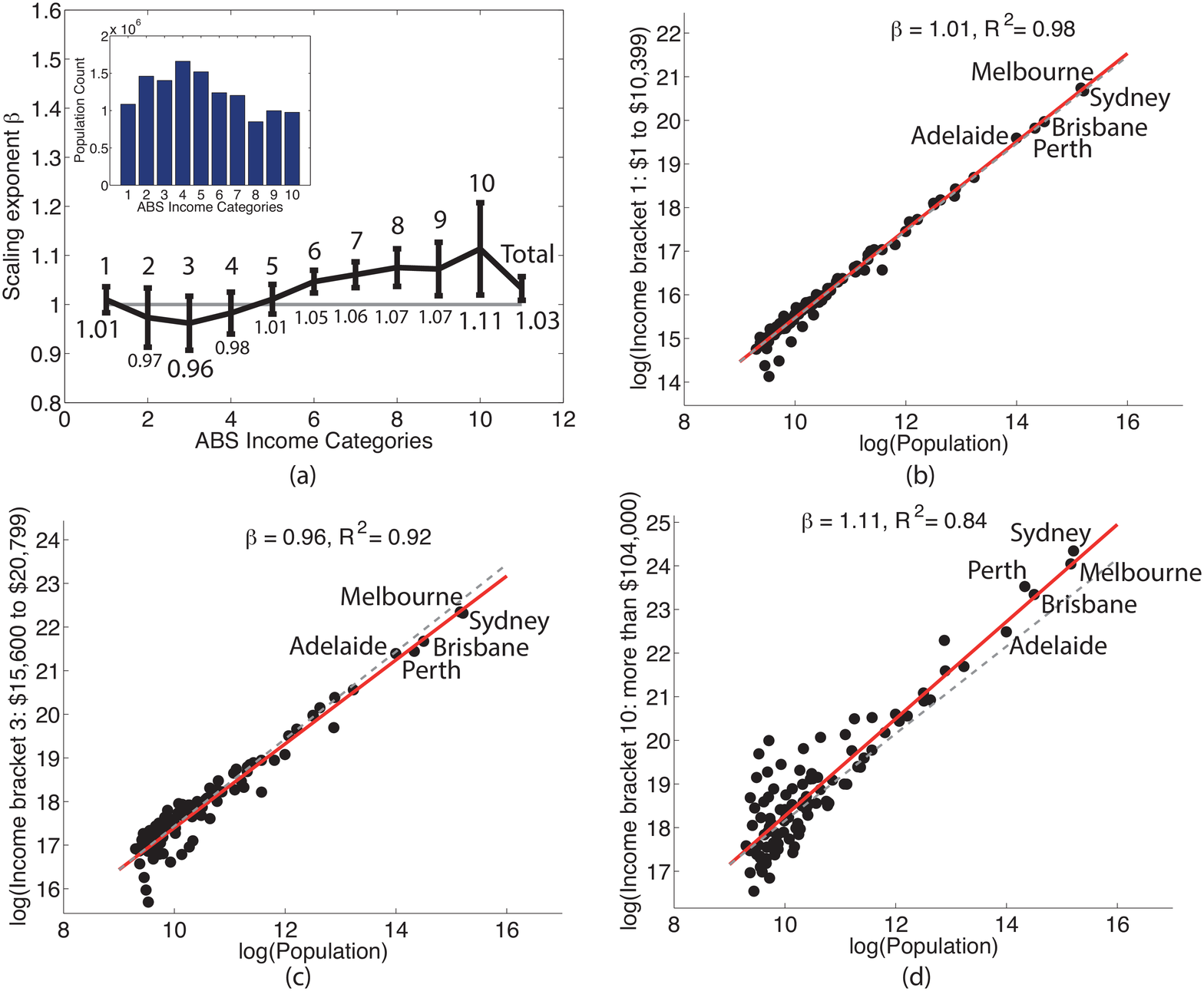}
\caption{Scaling of income in all 101 SUAs. (a) Scaling exponents with 95\% CI in 10 ABS income categories. (b-d) Scaling behavior for income categories 1, 3, and 10, (roughly categorised as lowest, middle lower, and highest, respectively), showing linear to sub-linear behaviors in the lower income categories, with superlinear behavior emerging for higher income categories.}
\label{Fig1}
\end{figure}

Figure~\ref{Fig1}(a) shows the scaling exponents for all the separate income categories and the total income for all the 101 SUAs. Figure~\ref{Fig1}(b-d) shows the scaling behavior for ABS income categories 1, 3, and 10, (roughly categorised as lowest, middle lower, and highest, respectively). Total income scales just superlinearly, with $\beta = 1.03$. This is different from the exponents previously reported, for the USA, European countries and China, where a clear superlinear behavior was reported, with higher $\beta$ values~\cite{bettencourt2007}. However, in more recent work for the UK, a similar close to linear scaling of total income has been reported~\cite{arcaute2015}. 

It is also observed that the lowest 5 income categories show sublinear or just linear exponents, and the top 5 higher income categories show superlinear exponents. Here we consider an exponent as sublinear, linear, or superlinear when the 95\% Confidence Interval (CI) error bar is safely below, at, around, or above the unity line, respectively. Thus, it is seen that the first 5 income categories show linear to sublinear behavior, whereas the highest 5 income categories show superlinear behavior, monotonically increasing especially for the top 5 income categories. Particularly interesting is the behavior of the highest income category, which is higher than all the others by the largest successive difference between exponents, Figure~\ref{Fig1}(d). A slightly larger spread or variance of distribution of the data points can also be observed for the highest income category, for some of the middle and low sized SUAs, ($R^{2} = 0.84$). 

This behavior implies that as cities grow bigger in population count, the incomes in the highest income categories grow (disproportionately) faster than incomes in the lower income categories. However, as the inset of actual population count distribution for the 101 SUAs in Fig.~\ref{Fig1}(a) shows, the lower income categories contain the largest sections of the population. This points to the emergence of inequality, since it shows that the incomes of the larger sections of the population are growing sublinearly, whereas the incomes of the smaller sections of the population are growing superlinearly. 

To confirm this analysis, however, a more in-depth analysis of the urban area definitions was needed. In recent work it has been shown that the manner in which urban areas are defined can affect the values of the scaling exponents~\cite{arcaute2015}. In addition, in Australia, not all the SUAs are equivalent, since the largest SUAs have populations exceeding 1-4 million and densities close to 1 000 persons per square km., whereas the smallest SUAs (some of them surrounded by regional or rural hinterlands), have populations exceeding just 10 000, and densities close to 21 persons per square km. Clearly, not all urban areas are equally ``urban". Since the ABS SUA definitions are based on the type of economic and social activities in an SUA, labour force and journey to work region definitions (see Methods), and because of the special manner in which population is heterogenously distributed across Australia, a deeper look into considering alternate urban area definitions was warranted. In particular, we wanted to study how the behavior of the scaling exponent changes if more ``urban" areas are considered, and less ``urban" areas are excluded from the analysis.  

\subsection*{Scaling of income in different income categories: Population counts and density cut-offs}

\begin{figure}[]
\centering
\includegraphics[width=0.8\linewidth]{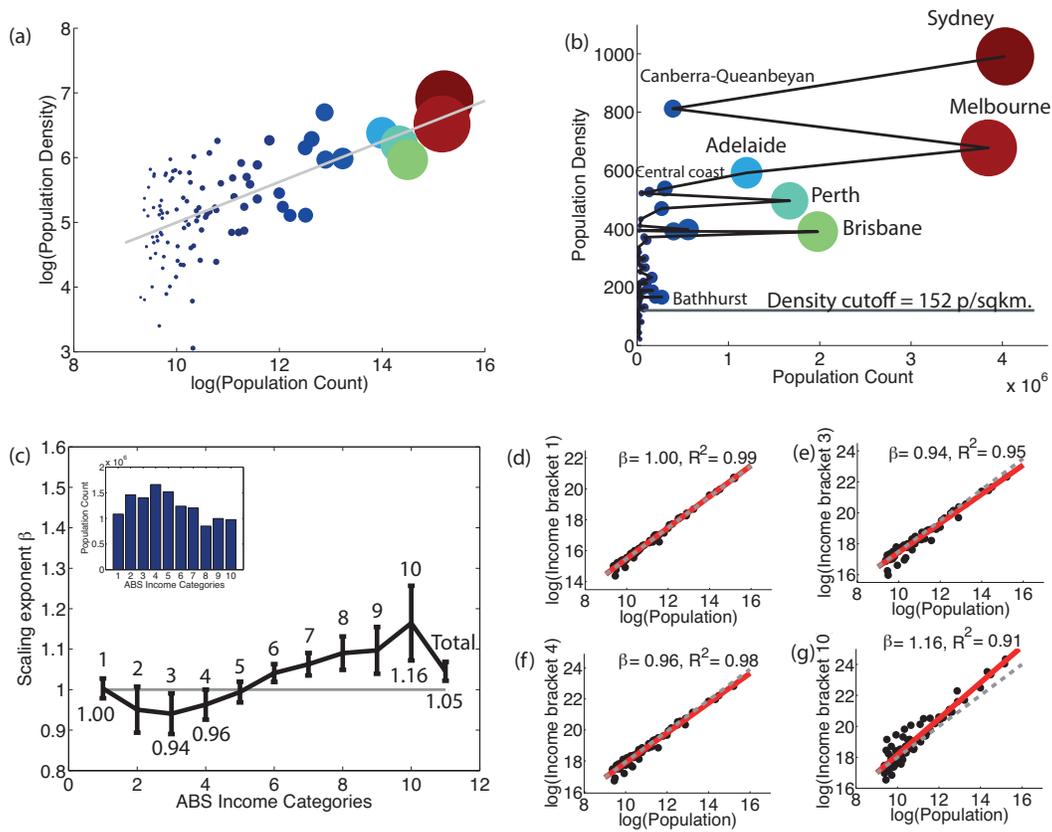}
\caption{Scaling of income in high population density SUAs. (a) Log-log plot of population count versus population density for all 101 SUAs shows the general positive correlation of higher population with higher density. Sizes of circles scaled corresponding to total population counts. (b) Plot of population count against population density shows some outliers. E.g. High population, lower densities, or low populations with higher densities. The cut-off point for excluding all SUAs where \textbf{both} population counts as well as population densities are low occurs at around 152 persons per sq km. (Bathhurst). (c) Scaling exponents with 95\% CI in 10 ABS income categories for the top 66 highest density and population count SUAs. (d-g) Scaling behavior for income categories 1, 3, 4, and 10, (roughly categorised as lowest, middle lower, and highest, respectively), showing linear to sub-linear behaviors in the lower income categories, with superlinear behavior emerging for higher income categories for the 66 high population and high density SUAs.}
\label{Fig2}
\end{figure}

In recent work, population density instead of population count has been proposed as one of the factors that can be used to define urban areas~\cite{arcaute2015}. Figure~\ref{Fig2}(a) shows the log-log plot of population count versus population densities (total population divided by total area in sq. km. reported by the ABS per SUA). A general positive correlation is observed with higher total population count areas showing higher population densities. However, as seen in Figure~\ref{Fig2}(b), a plot of total population counts against population densities also shows some outliers. For example, the Canberra-Queanbeyan region or Central Coast regions (clearly urban by all measures of socio-economic activity) have lower total populations than many larger total population count regions, but have higher density of population. Thus, we empirically identified a cut-off point using a population density of 152 persons per sq. km., below which all SUAs had low total population counts as well as low population density. This resulted in 66 SUAs that have either high total population count, or high population density, or both. We note again that as compared to most other countries of the world, these numbers on population density are particularly low. For example, the population density cut-off for the UK for the purpose of defining urban areas was 1400 persons per sq. km., but even the highest density SUA in Australia, Sydney, has a density of about 991 persons per sq. km.~\cite{arcaute2015}. Therefore, our identification of urban areas based on population density has been based on the relative comparitive positioning of the SUAs with each other. We note here that considering only the metropolitan areas will show higher population densities, for example Sydney and Melbourne metropolitans sit at about 1900 and 1500 persons per sq. km., respectively, but here we consider the entire SUA (Sydney and its cluster of related urban centres), which is more proper considering journey to work patterns, since people from urban areas outside the metropolitan region still travel daily to the city proper for work. 

Figure~\ref{Fig2}(c-g) show the behavior of the scaling exponents for these 66 SUAs. While the earlier reported pattern in Figure~\ref{Fig1} holds, the differences between the income categories becomes more pronounced, with the highest income category exponent going up to $\beta=1.16$, and the lowest one at $\beta=0.94$. Thus, considering higher population density and population areas show the inequality effect becoming more pronounced. 

\begin{figure}[]
\centering
\includegraphics[width=1.00\linewidth]{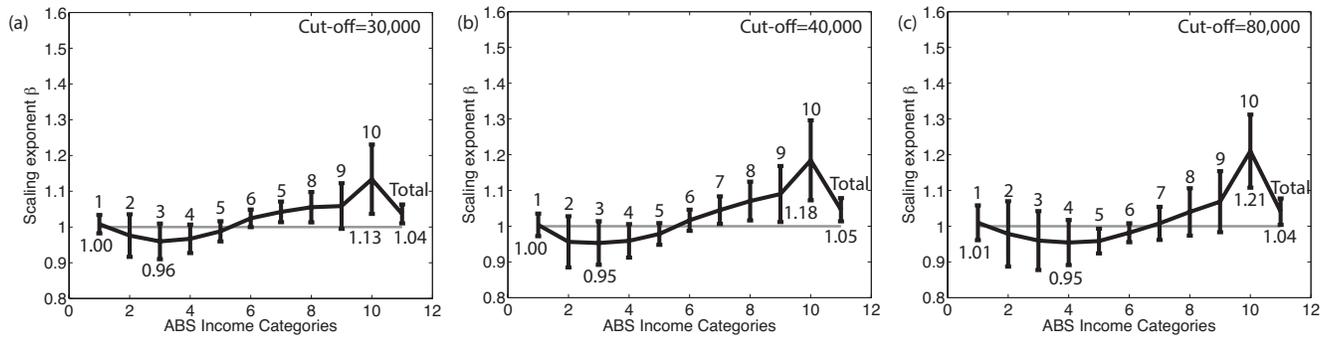}
\caption{Scaling of income in SUAs by population count cut-offs. Scaling exponents with 95\% CI in 10 ABS income categories for SUAs over (a) 30~000 total population, (b) 40~000 population, and (c) 80~000 population, showing linear to sub-linear behaviors in the lower income categories, with superlinear behavior emerging for higher income categories.}
\label{Fig3}
\end{figure}

Further, since the correlation between total population count and population density is strong, we also apply population count cut-offs, to consider urban areas larger than a total population count threshold. For example, there is only 1 SUA with a population of more than 80 000, that has a population density of 130 persons per sq. km. (that is below the density cut-off of 152 persons per sq. km.). Similarly, only 4 SUAs with a population of more than 40 000, and 8 SUAs with a population of more than 30 000, have a population density of lower than 152 persons per sq. km. Figure~\ref{Fig3}(a-c) shows the behavior of the scaling exponent for 45, 33, and 21 SUAs above the total population counts of 30~000, 40~000, and 80~000, respectively. Once again, the same trend is observed: as SUAs with larger populations and population densities are considered in urban area definitions, the scaling behavior shows the exponent for the highest income bracket rising more sharply, with the inequality effect becoming more pronounced. For example, when the SUAs beyond a total population count of 80 000 are considered, the exponent for the highest income category goes up to $\beta = 1.21$, whereas the lowest $\beta$ holds at $\beta = 0.95$. Further, it is interesting to note that some of the error bars for the other high income brackets actually graze or go below the unity line, showing the substantially different behavior of the top income bracket. However, the general trend of the monotonically rising exponent for higher income brackets is maintained.   

Thus, across all urban area definitions based on total population counts and population density, we find that lowest incomes grow just linearly or sublinearly ($\beta = 0.94$ to $1.01$), whereas highest incomes grow superlinearly ($\beta = 1.00$ to $1.21$), with total income just superlinear ($\beta = 1.03$ to $1.05$). These findings support the earlier finding: the bigger the city, the richer the city. But, we also see an emergent metric of inequality: the larger the population of the city, higher incomes grow more quickly than lower, suggesting a disproportionate agglomeration of wealth in the highest income categories in big cities. Because there are many more people on lower incomes that scale sublinearly as compared to the highest in almost all cities that scale superlinearly, these findings suggest agglomeration of income with a scaling of income inequality: inequalities become more pronounced with population size; larger the population, the greater is the inequality. 

Separately, we compute the Gini coefficient for the all the SUAs with more than a population of 30,000, using the standard definition of the Gini coefficient as the Lorenz curve, computing the proportion of the total income of the population that is cumulatively earned by the bottom x\% of the population. A Gini coefficient of 0.47 results by considering all the income categories (including the negative and nil income categories, for which we have the population counts, but no earnings). We have considered a zero earning for both the negative and nil earning groups, even though the imputed median for the negative earning categoy by the ABS is -\$101. Since we could not include these two categories in our scaling analysis, a Gini coefficient of 0.42 results by considering only the income categories reporting positive income, the ones we have used for our scaling analysis. As compared to the Australian average of 0.320 as reported by the ABS, and as compared to developing country averages, this is a high Gini coefficient. One reason why the official Gini coefficient of 0.320 is lower than 0.42 or 0.47, is that in our computations total income reported in the census is considered, whereas the Australian Gini of 0.320 is based on real income computations, and emerges lower than what our computations show. Nonetheless, equivalently, the Gini coefficient does signify high levels of income inequality in urban areas of the country. While the Gini coefficient looks at intra-population cumulative distributions of income, our findings relate the amount of inequality to the sizes of populations, and therefore, the sizes of cities or urban systems.




%
%

\section*{Discussion}

In this paper, we have examined the scaling of income in different income categories measured against population size as a measure of city size for the country of Australia. Australia represents a nascent urban system, still young and in early stages of growth, with a highly urbanized populace, but concentrated in very few urban areas, presenting a unique opportunity for this type of study. Using income and population data from the census of Australia, it is shown that there is a disproportionate agglomeration of income in the highest incomes categories in big cities. It was found under several urban area definitions based on total population counts and population density, that while incomes in the lower income groups scale sublinearly, the incomes in progressively higher income groups is superlinear. Moreover, the scaling of the income in the highest income category is significantly superlinear and high, as compared to all the other income categories. In general, there is a monotonically increasing trend for the scaling exponent for the high income brackets.  

This finding is a cause of concern for urban planning and economic policy, because it raises questions on equity, social justice and distributive justice aspects of big cities. Australia is not considered to be one of the most unequal nations on the planet. Recent studies have found that developing nations such as India and China, and developed nations such as the USA, are much more unequal than Australia~\cite{atkinson2015, piketty2014, wilkinson2009}. However, this paper shows scaling of income inequality for income data in a nation that is usually considered to be more equitable than many others. 

According to the ABS Household Income and Income Distribution of Australia (Cat. 6523.0, 2011-12), ``the wealthiest 20\% of households in Australia account for 61\% of total household net worth, with an average net worth of \$2.2 million per household, whereas the poorest 20\% of households account for just 1\% of total household net worth, with an average net worth of \$31,205 per household". Further, quoting again from the ABS Household Income and Income Distribution of Australia (Cat. 6523.0, 2011-12), ``while the mean equivalised disposable household income of all households in Australia in 2011-12 was \$918 per week, the median (i.e. the midpoint when all people are ranked in ascending order of income) was somewhat lower at \$790. This difference reflects the typically asymmetric distribution of income where a relatively small number of people have relatively very high household incomes, and a large number of people have relatively lower household incomes." 

Thus, from our findings it appears that the larger the city, the larger the growth of income in the highest income categories. This would be a good and prosperous scenario if there were a large number of people in the highest income categories. But, the ABS census statistics also shows, that nationally, there are few people in the highest income categories commanding a large portion of the income, with a very large number of people commanding a much smaller portion of the income. The scaling relationship in this paper, therefore, reveals, a scaling of inequality: inequalities seem to become more pronounced with population size; larger the population, greater was the inequality found, in Australia. It would be useful to test the scaling of income inequality on data from other countries around the world, especially countries that are reportedly more unequal than Australia.

Finally, computation of real income (as opposed to total reported income) in the income categories will be performed in our future work. For example, Australia has a social benefits system, where the population in the lower income brackets would be receiving income transfers and tax benefits, when their income is below a certain level. In such a scenario, the real incomes of the lower income brackets would be higher. However, because of the unique condition that a few urban areas house most of the population as well as all the economic opportunities, the costs of living (especially the largest proportion of expenditure, housing costs) are extremely high, most notably in Sydney and Melbourne. This leads to a situation where the poor spend a substantially high proportion of their income in housing and travel costs, whereas the rich spend, comparitively, a much lower proportion of their income for the same. Thus, from these types of factors, it is not immediately clear that the scaling of post-tax or any measure of real incomes will necessarily be more equally distributed than the scaling of pre-tax or gross measures of income. Computation of real or disposable income in the income brackets may change the scaling behavior reported in this paper, and it will be interesting to see whether real income computation intensifies the inequality observed in the gross case, or takes it towards increasing equity. Similarly, the scaling of other forms economic inequality, separate from income inequality, may also be quantified and studied in the same manner. Thus, the approach overall would be nonetheless useful in studying the scaling of income or other forms of economic inequality for a country.   

The findings are useful for informing urban planning and economic policy, particularly insights into how the size of an urban system may be an implicit driver for inequality, and therefore ask policy to examine ways in which resource distributions in larger cities can be made more equitable. 

\section*{Methods}

There is considerable debate in the literature on the geography that is adopted for the analysis of scaling behavior. As recent research shows, changing the geographical definitions can produce significant quantitative changes to the scaling exponent, and implied qualitative changes of interpreted urban dynamics~\cite{arcaute2015}. For example, a range of indicators have been reported for which the scaling exponent fluctuates between the superlinear and sublinear regimes, depending on the way in which the geography of analysis is defined~\cite{arcaute2015}, and specifically for comparisons of $CO_{2}$ emissions with city size, some studies have shown sub-linear regime relationships, while others have shown super-linear scaling relationships~\cite{fragkias2013, rybski2014, oliveira2014, louf2014}. Thus, any claim of scaling must rest on the behavior being tested over multiple possible realizations of reasonably defined geographies. Both the way in which data is measured and collected, as well as unlerlying factors other than city size have been proposed as reasons behind deviations, and therefore city and size and geography need careful examination.  

\subsection*{Australian Statistical Geography Standard}

For the Australian case, the country is divided into 8 States and Territories. The data for the entire country is collected and organized by a single national statistical body, the Australian Bureau of Statistics (ABS). The ABS defines the Australian Standard Geographic Classification (ASGC) that in July 2011 changed over to the new system of Australian Statistical Geography Standard (ASGS), that is a set of hierarchically organized levels of geographic units that correspond to spatial scales into which the entire country is progressively divided into, without overlaps and gaps~\cite{pink2011, asgs2011}. The areas are defined with regard to several important factors such as population cut-offs, social and economic activity, and labor force and housing markets and sub markets, along with ensuring the best possible consistency with administrative boundaries. However, the important fact to be noted is that the statistical bases for defining these geographies is not purely administrative, but several related factors as have been noted to be important in previous research~\cite{arcaute2015}. Australia is highly urbanized, with its urban population concentrated in very few large urban centres, making the derivation of the urban geography consistent in terms of many factors. Here we describe in brief the organization of the ASGC and the main spatial scales, and the details of how we have decided on a stable geography for analysis. 

The ABS classifies geographic structures into 2 classes: statistical ABS structures, and non-ABS political and administrative structures into which fall state suburbs (SSC), postal areas, and Local Government Areas (LGA). Here we consider the statistical ABS structures, since these are derived on the basis of social and economic interactions primarily concentrated within areas, rather than purely administrative divisions.  

The smallest geographic ABS structures are the \textbf{Statistical Areas Level 1 (SA1)}, with a minumum population of 200, and a maximum population of 800, with an average of approximately 400 people. From an administrative perspective, the SA1s closely reflect (though are not identical with) the non-ABS state suburbs and postal areas. The criteria for urban SA1s and rural SA1s are separately defined, with the urban SA1s characterized by presence of different types of developments such as airports, ports, large sports and educational campuses, roads, and large infrastructure. There are about 55 000 SA1s for the whole of Australia. From continuous aggreagtes of SA1s, \textbf{Urban Centres and Localities (UCLs)} are defined, with centres with core urban populations of more than 1 000 designated as \textbf{Urban Centres}, and centres with a core urban population of 200 to 1 000 designated as \textbf{Localities}. Populations contained within Urban Centres are used to describe Australia's urban population at the lowest geographic level. There are about 684 Urban Centres in Australia. 

We do not use the definition Urban Centres to define the geography of the city unit for our analysis, because several semi-urban and peripheral areas that are complete SA1s are predominantly surrounded by rural territory, making these the smallest level complete urban areas. On the other hand, many SA1s (for example, contiguous SA1s or UCLs within the Sydney, Melbourne and other capital cities regions) cannot be considered to be separate entities, since they are also parts of much larger urban agglomerations. Further, total reported income data for statistical areas is only available for the larger definitions of SA2, SA3, and SA4 (see below). 

The second level of geographic units are the \textbf{Statistical Areas Level 2 (SA2)}, with a minimum, maximum, and average population of $3 000$, $25 000$, and $10 000$, respectively. There are about 2~300 SA2s defined for the whole of Australia. Continuous SA1s are aggregated to produce each SA2, and the SA2 is designed to reflect functional areas, with the aim of representing a community that interacts together socially and economically. They also coincide largely with the non-ABS structure of Local Government Areas (LGA). 

From continuous aggregates of SA2s, \textbf{Significant Urban Areas (SUAs)} are defined, that describe extended urban concentrations of more than 10 000 people. That is, these SUAs do not represent a single Urban Centre, but they can represent a cluster of related Urban Centres, where the population interacts socially, lives, and travels for work. There are 101 SUAs describing urban concentrations in Australia. For reasons that we outline below, we adopt SUAs as the primary unit of defining geography in our work. 

The third level of geographic units are the \textbf{Statistical Areas Level 3 (SA3)}, with a minimum, maximum, and average population of approximately 30 000 to 130 000. They are designed as aggregates of whole SA2s, and reflect major regions within States and capture regional level outputs. No significant urban areas are defined for SA3s, and SA3s may include regional and rural areas.  

The fourth and largest level of geographic units are the \textbf{Statistical Areas Level 4 (SA4)}, with a minimum, maximum, and average population of approximately 100 000 to 500 000. In regional areas, SA4s contain populations of 100 000 to 300 000 whereas in metropolitan areas, SA4s have larger populations ranging from 300 000 to 500 000. They are designed as aggregates of whole SA3s. There are 106 SA4s in Australia. Whole SA4s aggreagte to \textbf{Greater Capital City Statistical Areas (GCCSA)} and \textbf{State and Territory}, with the GCCSAs focused on major urban cities and extending to the urban periphery of these large cities. 

SA4s are not adopted as the unit of defining geography in this paper, because while separate SA4s inside the Greater Capital City Statistical areas cannot be considered to be separate cities (for example, it would be wrong, artificial and arbitrary to consider the different SA4s inside the Sydney region as separate cities), while those that are outside the GCCSAs, can be considered to be smaller but complete urban areas or cities.  

For this work, we therefore work with the SUAs or Significant Urban Areas, that include all the Urban Centres within a region that interact closely socially and economically (including living and working within the same SUA). The reason for choosing SUAs is as follows: all major captial cities are defined as single urban clusters (for example, Sydney, Melbourne, Brisbane, Adelaide and Perth form the largest ones), whereas smaller urban areas surrounded by predominantly regional or rural areas, but that are clearly urban centres are defined as separate cities. SUA definitions therefore do not suffer from the limitation as outlined in SA4 and SA1 definitions: we are assured that no contiguous urban areas are arbitrarily defined as two cities just on account of an administrative or census division, and we are assured that identifiable urban areas of all sizes are identified nonetheless, even if they are surrounded by rural or regional hinterland, but are independently functioning urban regions. 
  
\subsection*{Computation of incomes in separate income categories}

For all the geographic area definitions discussed above, per capita weekly income statistics are available. The ABS has 10 per capita income categories, shown in Table~\ref{tab1}~\cite{abs1}. Categories (range identifiers) 01 and 02 represent negative and nil incomes, respectively, and the other income categories represent positive incomes.   

\begin{table}[h]
\centering
\small
\begin{tabular}{|l|l|l|l|} \hline
Range Identifier & Weekly Personal Income & Per annum personal income & Imputed Median   \\ \hline \hline
01 & Negative income & -- & \$ $-101$ \\ \hline
02 & Nil income & -- & \$ 0 \\ \hline
03 & \$1 - \$199 & \$1 - \$10,399 & \$80 \\ \hline
04 & \$200 - \$299 & \$10,400 - \$15,599 & \$263 \\ \hline
05 & \$300 - \$399 & \$15,600 - \$20,799 & \$349 \\ \hline
06 & \$400 - \$599 & \$20,800 - \$31,199 & \$487 \\ \hline
07 & \$600 - \$799 & \$31,200 - \$41,599 & \$698 \\ \hline
08 & \$800 - \$999 & \$41,600 - \$51,999 & \$896 \\ \hline
09 & \$1,000 - \$1,249 & \$52,000 - \$64,999 & \$1,107 \\ \hline
10 & \$1,250 - \$1,499 & \$65,000 - \$77,999 & \$1,363 \\ \hline
11 & \$1,500 - \$1,999 & \$78,000 - \$103,999 & \$1,695 \\ \hline
12 & \$2,000 or more & \$104,000 or more & \$2,579 \\ \hline
\end{tabular}
\caption{Personal Income Ranges and Imputed Medians}
\label{tab1}
\end{table}

Using the 2011 Census Income Data retrieved using the TableBuilder facility~\cite{abs2}, counts of the number of people in a particular income category on census night has been retrieved as per the SUA area definitions. That is, for each SUA, we have the number of persons in each of the above 10 income categories. We have computed the median earnings for each income category by multiplying the impiuted median income for each category for each SUA by the number of people in that income category. We call this the \textit{computed income} per income category. Adding across all income categories gives us the total annual computed income per SUA for the census year 2011. 

\subsection*{Analysis of scaling behavior}
For each SUA and for each income category, the log of population is plotted against the log of the computed income in that income caegory. Matlab's linear model fitting is used to calculate the $\beta$ exponent and the $R-$squared values. 


\section*{Acknowledgements}

The work in this paper has been supported by a University of Sydney Henry Halloran Trust Blue Sky Grant and a Research Incubator Fellowship. The authors thank Peter Robinson and Andy Dong for helpful discussions and comments.  

\section*{Author contributions statement}

SS conceived the experiment(s),  SS and RS conducted the experiment(s), SS, RS and PP analyzed the results, SW helped with data processing. All authors reviewed the manuscript. 

\section*{Additional information}

The authors declare no competing financial or other interests. 





%


\end{document}